\documentclass[a4paper,11pt]{article}
\usepackage{float}
\usepackage[pdftex]{graphicx}
\usepackage{subcaption}
\usepackage[T1]{fontenc}
\usepackage{lmodern}
\usepackage{slashed}
\usepackage[utf8]{inputenc}
\usepackage[english]{babel}
\usepackage{microtype}
\usepackage{cite}
\usepackage{amsmath,amssymb,amsfonts,amsthm}
\usepackage{mathtools,mathrsfs,calligra,aurical}
\usepackage[nottoc,notlot,notlof]{tocbibind}
\usepackage{upgreek}
\usepackage{mathtools}
\allowdisplaybreaks
\usepackage[all]{xy}
\usepackage{color} 
\usepackage{xcolor}
\usepackage{graphicx}
\usepackage{cancel}
\graphicspath{{images/}}
\usepackage{geometry}
\usepackage[toc,page]{appendix}
\usepackage{hyperref}
\usepackage[normalem]{ulem}

\usepackage{bm}
\usepackage{ragged2e}
\usepackage{appendix}
\usepackage{slashed}
\usepackage{bbold}
\usepackage{cancel}

\definecolor{blue-violet}{rgb}{0.54, 0.17, 0.89}
\definecolor{PineGreen}{cmyk}{0.92, 0, 0.59, 0.25}
\definecolor{YellowOrange}{cmyk}{0, 0.42, 1, 0}
\definecolor{orange}{rgb}{0.95, 0.5, 0.1}

\newcommand{\be}{\begin{equation}}
\newcommand{\bea}{\begin{eqnarray}}

\newcommand{\ee}{\end{equation}}
\newcommand{\eea}{\end{eqnarray}}

\DeclareMathAlphabet{\mathpzc}{OT1}{pzc}{m}{it}

\usepackage[dvipsnames]{xcolor}

\begin{document}

\begin{titlepage}
\begin{flushright}
\par\end{flushright}
\vskip 0.5cm
\begin{center}
\textbf{\LARGE \bf  Equation of State of a Strongly Coupled Perfect Fluid with Spin from Holography}\\
\vskip 5mm

\vskip 1cm

\large {\bf Andr\'{e}s Anabal\'{o}n}$^{~a ~b}$\footnote{anabalo@gmail.com}  \large {and \bf Horatiu Nastase}$^{~b}$\footnote{horatiu.nastase@unesp.br}

\vskip .5cm 

$^{(a)}${\textit{Departamento de F\'isica, Universidad de Concepci\'on, Casilla 160-C, Concepci\'on, Chile.}}\\ \vskip .1cm
$^{(b)}${\textit{Instituto de F\'isica Te\'orica, UNESP-Universidade Estadual Paulista \\
R. Dr. Bento T. Ferraz 271, Bl. II, Sao Paulo 01140-070, SP, Brazil.}}
\end{center}
\begin{abstract}
We derive the equation of state of a strongly coupled relativistic perfect fluid with finite spin in $2+1$ dimensions using holography. The dual gravitational description is provided by a spinning black hole in $AdS_4$, whose Hawking temperature, angular velocity, and Bekenstein-Hawking entropy determine the thermodynamic properties of the boundary fluid. We obtain the equation of state and analyze its local thermodynamic stability, finding a critical rotation above which the fluid becomes thermodynamically unstable providing a bound for the rotation of a strongly coupled quark gluon plasma at temperature $T$ and angular velocity $\omega$ given by $\frac{4\pi T}{\omega}> 2.77239$. Remarkably, the same gravitational solution contains a second black hole associated with a second boundary. Although this black hole is non-spinning, its dual fluid rotates and possesses a distinct, ``exotic'' equation of state. Our results provide a holographic equation of state for strongly coupled matter with finite spin and establish a direct connection between black-hole rotation, intrinsic angular momentum, and the thermodynamic stability of relativistic fluids.
\end{abstract}

\vfill{}
\vspace{1.5cm}
\end{titlepage}

\setcounter{footnote}{0}

Relativistic heavy-ion collisions provide an experimental realization of strongly interacting matter that behaves, to a remarkable degree, as a nearly perfect fluid. Non-central collisions carry enormous angular momentum, generating a strongly vortical state of the quark-gluon plasma. The STAR Collaboration established experimentally that the spin of emitted $\Lambda$ and $\bar{\Lambda}$ hyperons is correlated with the angular momentum of the collision, providing direct evidence that the vorticity of the fluid couples to particle spin \cite{STAR:2017ckg}. Spin density has therefore become an experimentally accessible property of strongly interacting matter.

These observations motivate an extension of the equilibrium thermodynamics of perfect fluids in which intrinsic angular momentum is treated as an independent thermodynamic variable. Relativistic spin hydrodynamics introduces a tensor spin chemical potential conjugate to the spin angular-momentum density, providing a natural framework for describing such states \cite{Florkowski:2018fap}. The corresponding equilibrium theory requires an equation of state relating the pressure and energy density to the temperature and spin density.

While considerable progress has been made in formulating the dynamics of spin-polarized fluids, no equation of state of an equilibrium strongly interacting perfect fluid with finite spin is known, although there have been some claims \cite{Florkowski:2024bfw,Armas:2026bmw}.\footnote{Note that in standard holography, for 
instance \cite{Hawking:1998kw,Papadimitriou:2005ii,Braga:2025wox}, one 
consides just orbital angular momentum coming from the bulk.}
The aim of this work is to provide the first such equation of state in $2+1$ dimensions and explore its immediate thermodynamic consequences.

The equation of state is derived from our recent discovery \cite{toappear} of a spinning black hole in $AdS_4$, 
see the appendix, characterized by Hawking temperature, angular velocity, and Bekenstein-Hawking entropy density
\begin{equation}\label{Hawk}
T_{BH}=
\frac{r_+\,\Omega_0\left(3r_+^2+1\right)}{4\pi\left(1+r_+^2\right)}\, , \qquad \Omega_{BH}=\frac{\Omega_0 r_+^2}{1+r_+^2}\, , \qquad s_{BH}=\frac{L^2}{\kappa}2\pi\,\Omega_0^2\left(1+r_+^2\right) \gamma^3
\end{equation}
here $\gamma=(1-w^2\Omega_0^2)^{-1/2}$ is the usual Lorentz factor. These quantities are measured with respect to the static (Laboratory) frame
\begin{equation}
ds^2=-dt^2+dw^2+w^2d\phi^2\, .
\end{equation}
$r_+$ is the location of the horizon of the black hole. It is otherwise an implicit function of $T_{BH}/\Omega_0$, determined by \eqref{Hawk}. The black hole is dual to a conformal fluid rotating with angular velocity $\Omega_0$. The gravitational origin of these thermodynamic variables can be traced back to the overall factor appearing in the area law that includes the $AdS_4$ radius $L$ and the reduced Newton constant $\kappa$, which, in terms of gauge theory variables, is $L^2/\kappa=\frac{\sqrt{2}}{12\pi}k^{1/2}N^{3/2}$, where $k$ and $N$ are the levels and ranks of the gauge groups, respectively, of a three dimensional Chern-Simons theory known as the ABJ(M) theory, dual to the $AdS_4$ gravity.

The boundary energy density $\rho_E$ and 
{\em orbital} angular momentum density $j$ in the inertial frame
are 
\begin{equation}
\rho_E=T_{tt}=\frac{
L^2}{2\kappa}\Omega_0^3 r_+ \left(r_+^2+1\right)
\left(2+w^2\Omega_0^2\right)
\gamma^5\, ,\quad
j=T{^t}_{\phi}=3\frac{
L^2}{2\kappa}\Omega_0^4\,w^2\,r_+
\left(r_+^2+1\right)
\gamma^5\, ,
\end{equation}
but in the inertial frame 
$\rho_E=\vec{\Omega_0}\cdot \vec{j}_T
=\Omega_0j_T$,\footnote{This formula is usually defined in the presence of only orbital angular momentum, 
see for instance \cite{Guemez}, but we here extend it to a total angular momentum.} so we have also an extra density associated with 
intrinsic spin, $\sigma=j_T-j=\rho_E/\Omega_0-j$, which in the 
comoving frame becomes 
\begin{equation}
\sigma_{\rm com}=\sigma \gamma^{-1}
=2\frac{L^2}{2\kappa}r_+(r_+^2+1)
\gamma^2\Omega_0.
\end{equation}

The pressure and energy density of the fluid are, in the 
co-moving frame (such that 
$\langle T_{\mu\nu}\rangle=(\rho+P)U_\mu U_\nu+Pg_{\mu\nu}$),
\begin{equation}
P=\frac{\rho}{2}=\frac{L^2}{\kappa}\frac{r_+ \Omega_0^3 \left(1+r_+^2\right)}{2}\gamma^3\, .
\end{equation}
In the co-moving frame the angular velocity of the black hole\footnote{Indeed, $\Omega$ is minus the angular velocity of the black hole, measured in the co-moving frame. $\Omega$ is positive and makes the results simpler to express. We remark that our calculation of $\Omega_{BH}$ is done in a system coordinates that is not spinning at infinity $\Omega_{\infty}=0$.} and its temperature are\footnote{Note that $T_{BH}$ is a 
temperature associated with the {\em whole} spacetime, through 
the periodicity in Euclidean signature, which is why it is a QFT
temperature of the static (inertial) frame, and the comoving 
frame rotates with respect to it.}
\begin{equation}
T=\gamma T_{BH} \, , \qquad \Omega=\gamma (\Omega_0-\Omega_{BH})=\gamma\frac{\Omega_0}{1+r_+^2} \, .
\end{equation}
The extensive variables in the co-moving frame are 
(the comoving frame is the proper frame)
\begin{equation}
P=\frac{\rho}{2}=\frac{L^2 r_+ \Omega^3 \left(1+r_+^2\right)^4}{2\kappa}\, ,\qquad \hat{s}_{BH}=\frac{2L^2\pi\,\Omega^2\left(1+r_+^2\right)^3}{\kappa}=s_{BH}\gamma^{-1}\, .
\end{equation}
In the co-moving frame there is no orbital angular momentum, 
and there is only the spin density 
$\sigma_{\rm com}$. 
The thermodynamical variables satisfy an extended Euler equation, 
\begin{equation}
\rho+P=T\hat{s}_{BH}+\Omega \sigma_{\rm com}
\, . 
\end{equation}
This is then consistent with the interpretation
in \cite{Becattini:2009wh} of the quantity multiplying 
$\Omega$ as the (intrinsic) spin density, which can be rewritten
in terms of black hole quantities as
\footnote{Note that for $T/\Omega_0
\rightarrow \infty$, thus for $\Omega_0\rightarrow 0$, giving $r_+\rightarrow \infty$, we get 
$T/\Omega_0=3r_+/(4\pi)$ and 
$\rho=\sigma_{\rm com}\Omega_0=\frac{2}{3}T\hat S_{BH}$, yet $\Omega \sigma_{\rm com}/\rho\rightarrow 0$, so both the thermodynamics and the Euler equation reduce to the standard ones.}
\begin{equation}
\sigma_{\rm com}=\frac{r_+}{2\pi} \hat{s}_{BH}
\end{equation}
Furthermore, given this identification, the standard thermodynamic relations are satisfied
\begin{equation}
\frac{\partial P}{\partial T} =\hat{s}_{BH}\, , \qquad \frac{\partial P}{\partial \Omega} =\sigma_{
\rm com} \, .
\end{equation}
The first law of thermodynamics follows from this fact. Indeed, the thermodynamic treatment of fluids with spin considers that the relevant variable is the Lorentz dilated angular velocity of the fluid \cite{Becattini:2009wh}, 
\begin{equation}
\omega=\gamma \Omega_0 \, .
\end{equation}
If we define $P=P(T,\omega)\equiv-G_{\omega}$ we get 
\begin{equation}
\frac{\partial P}{\partial T}=\hat s_{BH}+
\sigma_{\rm com}
\frac{\partial \Omega}{\partial T}\equiv s_{0}\, ,\qquad \frac{\partial P}{\partial \omega}=\sigma_{\rm com}\frac{\partial \Omega}{\partial \omega}\equiv \sigma_{0}\, ,
\end{equation}
where $s_0$ and $\sigma_0$ are the entropy density of the fluid and the spin density of the fluid respectively. $s_0$ is a positive definite function and is always smaller than $\hat{s}_{BH}$,
\begin{equation}
s_0=\frac{2\pi L^2  \omega^2 \left(3r_+^2+1\right)\left(r_+^2+1\right)^2}{\kappa\left(3r_+^4+8r_+^2+1\right)}\leq \hat{s}_{BH} \, ,
\end{equation}
Equality is attained only in the limits $r_+=0$ or $r_+\to\infty$, corresponding respectively to the absence of a thermal system or to the conformal fluid without spin limit. The determinant of the Hessian of the free energy yields
\begin{equation}
\frac{\partial^2 G_{\omega}}{\partial T^2}\,
\frac{\partial^2 G_{\omega}}{\partial \omega^2}
-
\left(
\frac{\partial^2 G_{\omega}}{\partial T\,\partial \omega}
\right)^2
=\frac{16\pi^2 L^4 \omega^2 \left(r_+^2+1\right)^4 \left(27r_+^6-27r_+^4-23r_+^2-1\right)}{\kappa^2\left(3r_+^4+8r_+^2+1\right)^3}
\;,
\end{equation}
and we find that the spinning fluid is locally stable for\footnote{By Sylvester's criterion, the Hessian is negative definite if its determinant is positive and the specific heat is positive. The latter condition is automatically satisfied for $r_+>r_*$.} $r_+>r_*$ ,
\begin{equation}
r_*\approx 1.24937\implies \frac{4\pi T}{\omega}> 2.77239\, .
\end{equation}
Thus, sufficiently rapid rotation drives the fluid into a thermodynamically unstable regime, potentially signaling the onset of a new phase, possibly associated with hadronization.

Remarkably, the gravitational solution contains a second black hole associated with a second boundary. Although this black hole itself is non-spinning, it is dual to a fluid rotating with angular velocity $\Omega_0$. In the co-moving frame of the fluid, its thermodynamic variables are\footnote{This fluid has a negative energy density and negative pressure as a consequence of the orientation of the spacetime. Here we provide a consistent thermodynamic formulation in which the physical energy density and pressure are positive.
}
\begin{equation}
P^E=\frac{\rho^E}{2}=\frac{L^2}{\kappa} 2\pi T\omega^2\, ,\qquad \hat{s}^E_{BH}= \frac{L^2}{\kappa}2\pi\omega^2\, ,\qquad \sigma^E = \frac{L^2}{\kappa} 4 \pi T \omega
\end{equation}
We denote the thermodynamic quantities of this fluid by the label ``E'', for exotic. Its thermodynamics is consistent in the sense that the first law of thermodynamics is satisfied, with the entropy of the fluid given by the Bekenstein-Hawking area-law entropy of the black hole. Since this entropy is independent of the temperature, the specific heat of the fluid vanishes. Consequently, the determinant of the Hessian of the free energy is negative, signaling thermodynamic instability, at least in the absence of a chemical potential or an external magnetic field.

\section*{Acknowledgements}
The work of HN is supported in part by  CNPq grant 
304583/2023-5 and FAPESP grant 2024/15298-0.
HN would also like to thank the ICTP-SAIFR for their support 
through FAPESP grant 2021/14335-0. The work of AA is supported in part by the FONDECYT grants 1230853, 1242043, 1250133, 1262452 and 1262414 and by the FAPESP grant 2024/16864-9. 

\appendix
\section{Black hole solution}

The solution we have found in \cite{toappear} will be 
considered in the absence of charge, so for $q=0$, which is 
\begin{eqnarray}
ds^2& =&\alpha^2
\left(\frac{
y^4 r m}{y^2+r^2}
-\frac{y^2 r^2}{L^2}
\right)dt^2
-\frac{2 \alpha y^2 r (y^2-1)
mL^2}{y^2+r^2}
\,dt\,d\phi\cr
&&+\left(\frac{r (y^2-1)^2mL^4}{y^2+r^2}
+L^2 (y^2-1)(r^2+1)
\right)d\phi^2
+\frac{L^2(y^2+r^2)}{y^2 (y^2-1)}dy^2
+\frac{y^2+r^2}{\frac{r^4}{L^2}+\frac{r^2}{L^2}-m r}
dr^2.\cr&&
\end{eqnarray}
In this case we have $m=r_+(r_+^2+1)/L^2$.


\hypersetup{linkcolor=blue}
\phantomsection
\addtocontents{toc}{\protect\addvspace{4.5pt}}

\bibliographystyle{mybibstyle}
\bibliography{bibliografia}

\providecommand{\href}[2]{#2}\begingroup\begin{thebibliography}{10}

\bibitem{STAR:2017ckg}
\textbf{STAR} Collaboration, L.~Adamczyk et~al., \textit{``{Global $\Lambda$ hyperon polarization in nuclear collisions: evidence for the most vortical fluid}''}, Nature \textbf{548} (2017) 62--65, [\href{http://arxiv.org/abs/1701.06657}{\texttt{arXiv:1701.06657}}].

\bibitem{Florkowski:2018fap}
W.~Florkowski, A.~Kumar and R.~Ryblewski, \textit{``{Relativistic hydrodynamics for spin-polarized fluids}''}, Prog. Part. Nucl. Phys. \textbf{108} (2019) 103709, [\href{http://arxiv.org/abs/1811.04409}{\texttt{arXiv:1811.04409}}].

\bibitem{Florkowski:2024bfw}
W.~Florkowski and M.~Hontarenko, \textit{``{Generalized Thermodynamic Relations for Perfect Spin Hydrodynamics}''}, Phys. Rev. Lett. \textbf{134} (2025), n.~8, 082302, [\href{http://arxiv.org/abs/2405.03263}{\texttt{arXiv:2405.03263}}].

\bibitem{Armas:2026bmw}
J.~Armas and A.~Jain, \textit{``{Thermodynamics of ideal spin fluids and pseudo-gauge ambiguity}''}, \href{http://arxiv.org/abs/2601.14421}{\texttt{arXiv:2601.14421}}.

\bibitem{Hawking:1998kw}
S.W. Hawking, C.J. Hunter and M.~Taylor, \textit{``{Rotation and the AdS / CFT correspondence}''}, Phys. Rev. D \textbf{59} (1999) 064005, [\href{http://arxiv.org/abs/hep-th/9811056}{\texttt{hep-th/9811056}}].

\bibitem{Papadimitriou:2005ii}
I.~Papadimitriou and K.~Skenderis, \textit{``{Thermodynamics of asymptotically locally AdS spacetimes}''}, JHEP \textbf{08} (2005) 004, [\href{http://arxiv.org/abs/hep-th/0505190}{\texttt{hep-th/0505190}}].

\bibitem{Braga:2025wox}
N.R.F. Braga and A.L. Ferreira, Jr., \textit{``{Unraveling the effect of rotation on the confinement/deconfinement transition of the quark-gluon plasma}''}, Phys. Rev. D \textbf{114} (2026), n.~1, 014029, [\href{http://arxiv.org/abs/2511.22464}{\texttt{arXiv:2511.22464}}].

\bibitem{toappear}
A.~Anabalon and H.~Nastase, \textit{``to appear''},.

\bibitem{Guemez}
J.~Guemez, M.~Fiolhais and L.~Fernandez, \textit{``Relativistic rotation - how does the energy vary with angular momentum?''}, J. Phys.: Conf. Ser. \textbf{1141} (2018) 012131.

\bibitem{Becattini:2009wh}
F.~Becattini and L.~Tinti, \textit{``{The Ideal relativistic rotating gas as a perfect fluid with spin}''}, Annals Phys. \textbf{325} (2010) 1566--1594, [\href{http://arxiv.org/abs/0911.0864}{\texttt{arXiv:0911.0864}}].

\end{thebibliography}\endgroup
\end{document}